\pdfoutput=1
\documentclass[11pt]{article}

\usepackage[T1]{fontenc}
\usepackage[utf8]{inputenc}
\usepackage{textcomp}
\usepackage{mathptmx}
\usepackage[scaled=0.92]{helvet}
\usepackage[margin=1in]{geometry}
\usepackage{graphicx}
\usepackage{float}
\usepackage[super,sort&compress,comma]{natbib}
\usepackage[font=small,labelfont=bf,labelsep=period]{caption}
\usepackage{microtype}
\usepackage[hidelinks]{hyperref}
\hypersetup{pdftitle={A software tool for parametric collimator creation for linac-based preclinical irradiation studies},pdfauthor={Joseph B. Schulz, Dixin Chen, Lloyd Emmanuel Kamole Ghomsi, M. Ramish Ashraf, Lawrie Skinner, Edward Elliot Graves, Billy W. Loo Jr, Stavros Melemenidis}}

\newcommand{\cm}[1]{#1~cm$^2$}
\newcommand{\by}{~$\times$~}

\title{\textbf{A software tool for parametric collimator creation for linac-based preclinical irradiation studies}}

\author{%
Joseph B. Schulz\textsuperscript{1,2,*},
Dixin Chen\textsuperscript{1,*},
Lloyd Emmanuel Kamole Ghomsi\textsuperscript{1},\\
M. Ramish Ashraf\textsuperscript{1},
Lawrie Skinner\textsuperscript{1},
Edward Elliot Graves\textsuperscript{1},\\
Billy W. Loo Jr\textsuperscript{1,\textdagger},
Stavros Melemenidis\textsuperscript{1,3,\textdagger}
\\[1em]
{\small \textsuperscript{1}Department of Radiation Oncology, Stanford University School of Medicine, Stanford, CA 94305, USA}\\
{\small \textsuperscript{2}Department of Medical Physics, University of Wisconsin--Madison, Madison, WI 53705, USA}\\
{\small \textsuperscript{3}Department of Radiation Oncology, University of Colorado Anschutz Medical Campus, Aurora, CO 80045, USA}\\[0.8em]
{\small \textsuperscript{*}Co-first authors.\quad \textsuperscript{\textdagger}Co-senior authors.}\\
{\small Corresponding author: Joseph B. Schulz, \href{mailto:joey.schulz@wisc.edu}{joey.schulz@wisc.edu}}
}

\date{}

\begin{document}

\maketitle

\begin{abstract}
\noindent\textbf{Background:} Designing custom Cerrobend collimators for linac-based preclinical electron irradiation can be labor-intensive and susceptible to design inconsistencies. A parametric workflow was developed and validated to design and fabricate a beam degrader, collimator mold, and collimator shell for linac-based small-animal irradiation, including organ-specific fields. The approach integrates computer-aided design and 3D printing so that geometry and jig alignment relative to the central axis can be specified directly from user inputs.

\noindent\textbf{Purpose:} To develop and validate an automated open-source parametric workflow for the design and fabrication of multi-part collimators for linac-based preclinical electron radiotherapy. This study evaluates geometric agreement and downstream beam profiles under conventional and ultra-high dose-rate (UHDR) electron delivery conditions.

\noindent\textbf{Methods:} From aperture and offset inputs, the tool exports STL models for a collimator mold, collimator shell, Cerrobend holder, and beam degrader. Two designs, a \cm{4.0\by1.5} bilateral lung collimator centered on the central axis and a \cm{1.5\by1.5} unilateral lung collimator with a 7.5~mm jig offset, both featuring a 15~mm thick Cerrobend collimator and a 10~mm PLA degrader, were fabricated and dosimetrically characterized with films under previously characterized conventional and UHDR delivery conditions. Kilovoltage planar images were taken to assess the geometry and uniformity of the fabricated collimators. Full width at half maximum (FWHM), penumbra, and off-axis leakage were measured.

\noindent\textbf{Results:} The physical Cerrobend apertures deviated by less than 0.2~mm from their specified designs. FWHM values were within 3\% of the nominal field dimensions with penumbra ranging from 0.73 to 1.05~mm under conventional and UHDR delivery. Crossline off-axis leakage doses at 2~cm from the collimator edge were below 2\% under conventional dose rate for both collimators.

\noindent\textbf{Conclusions:} Parametric collimator generation standardized digital design inputs for the two evaluated rectangular linac-based electron fields. Further work will assess specimen-level dosimetry, additional aperture geometries, and Monte Carlo integration.
\end{abstract}

\section{Introduction}

Small-animal irradiation studies underpin radiobiological research, guide treatment-schedule optimization and support evaluation of emerging modalities such as ultra-high dose-rate (UHDR) radiotherapy\cite{Hall2006,Kirsch2018,Verhaegen2011,Brown2022,Pizarro2017,Schuler2022}. Preclinical irradiation technologies include fixed or broad-beam cabinet x-ray systems and image-guided small-animal irradiators, including SARRP with cone-beam CT guidance\cite{Verhaegen2011,Brown2022,Bazalova2014,Graves2007,Wong2008}. Modified clinical linacs also provide electrons for preclinical irradiation, including UHDR workflows with fixed source-to-surface distance and simple collimation systems\cite{Schuler2017,Lempart2019,Jaccard2018}. Precise field shaping is essential because millimeter-scale misalignment of a collimator can alter dose distributions and confound experimental endpoints that can lead to translation of research to humans\cite{Verhaegen2018,Hackam2006}. Field shaping is platform-specific: cabinet and image-guided systems may use fixed apertures, jaws, or leaf blockers\cite{Verhaegen2011,Graves2007,Wong2008,Redler2021}, whereas linac-based electron workflows may use Cerrobend cutouts\cite{Schuler2017,Byrne2024}. Fabricating study-specific Cerrobend collimators is labor-intensive and susceptible to geometric error. Because many UHDR-capable linac beamlines lack dedicated field-shaping accessories, investigators rely on bespoke Cerrobend blocks\cite{Schuler2017,Lempart2019,Byrne2024,No2023}. This dependence on custom fabrication adds to broader reproducibility concerns in radiobiological research, where Draeger et al.\ identified incomplete physics and dosimetry reporting as a major barrier\cite{Draeger2020}.

Additive manufacturing and computer-aided design (CAD) now offer sub-millimeter accuracy and rapid prototyping for radiotherapy accessories, including collimators and compensators\cite{Rooney2020,McGarry2025,Skinner2019,Schulz2023,Schulz2024,Capaldi2024}. Parametric solid modeling allows aperture shape, stacking dimensions, and alignment of the specimen relative to the beam to be captured numerically and shared as print-ready meshes. This is useful when custom or study-specific collimators must be fabricated repeatedly and validated digital templates can also be reused when appropriate. The present work details a pipeline that outputs STL files for a stacked electron collimator composed of a Polylactic Acid (PLA) beam degrader and collimator shell, a Polyethylene Terephthalate Glycol (PETG) collimator mold, and a separate PLA Cerrobend holder, and validates beam-shaping performance on a Varian TrueBeam linac using Gafchromic EBT-XD radiochromic film (Ashland, Wayne, NJ) under conventional and UHDR electron beams. The software is open source, however, dosimetric validation in this study is limited to two rectangular lung fields. The linac configuration and beam-output characterization were reported previously\cite{Melemenidis2026}, and the present study focuses on CC and the resulting field-shaping hardware.

\section{Materials and Methods}

\subsection{Design Pipeline and Physical Collimator Assembly}

\begin{figure}[H]
  \centering
  \includegraphics[width=0.86\textwidth]{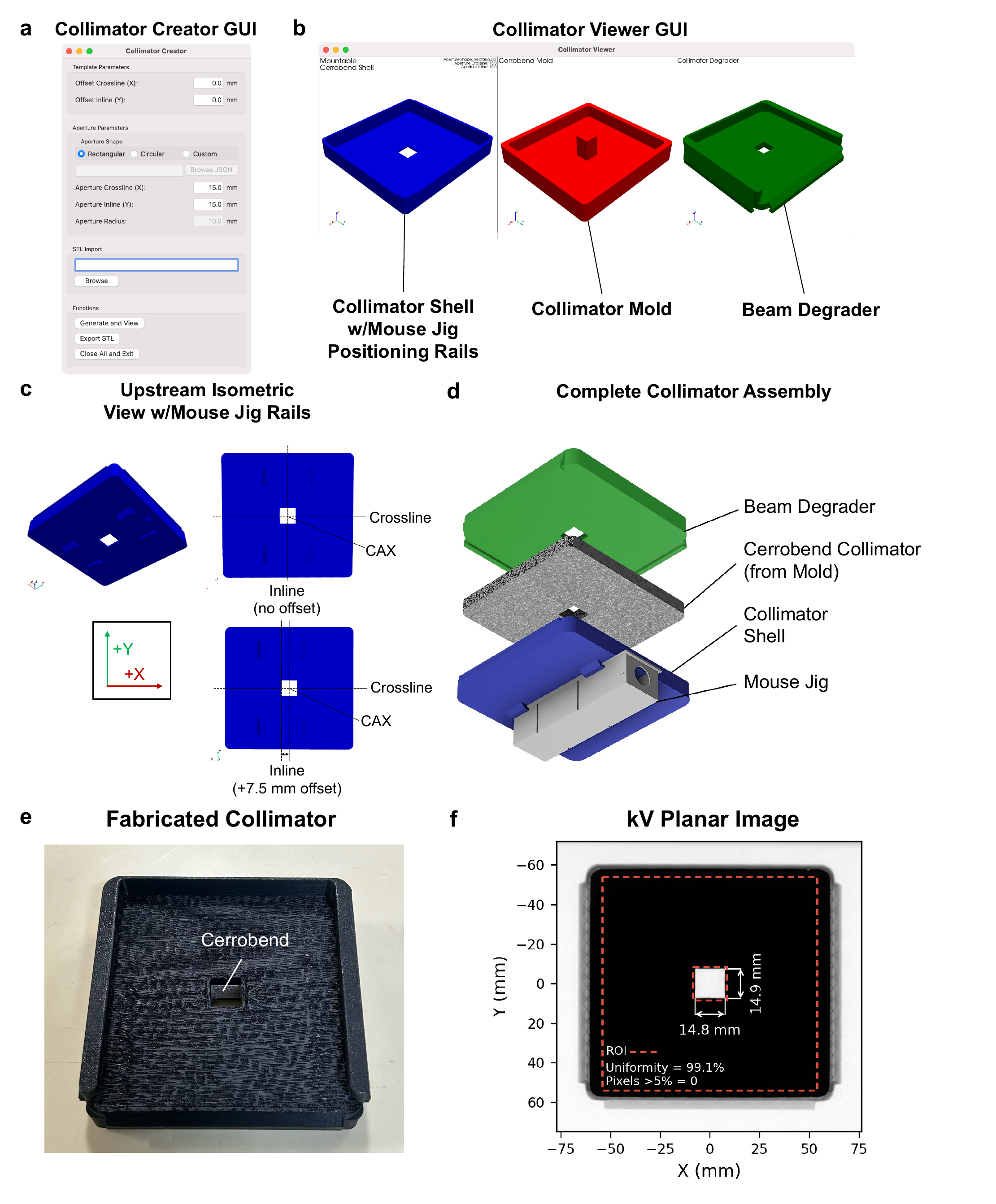}
  \caption{Collimator Creator software and modular collimator hardware. (\textbf{a}) Interface with template offsets (crossline X, inline Y), aperture shape and dimensions, optional STL import, and export controls. (\textbf{b}) Preview of generated solids: collimator shell with mouse-jig positioning rails, PETG collimator mold, and PLA beam degrader. (\textbf{c}) Collimator-shell underside and laboratory X/Y axes; beam's-eye depiction of the aperture on CAX with no inline offset versus a +7.5~mm inline offset. (\textbf{d}) Exploded stack: beam degrader, Cerrobend collimator after casting, collimator shell, and mouse jig slid onto the shell rails. (\textbf{e}) Photograph of the fabricated collimator shell with the Cerrobend collimator seated in the aperture with 7.5~mm offset. (\textbf{f}) Kilovoltage (kV) planar image of the unilateral lung collimator showing the 14.8\by14.9~mm aperture with the marked region of interest (ROI) where the uniformity was 99.1\% with 0 pixels deviating more than 5\% from the mean.}
  \label{fig:cc}
\end{figure}

\begin{figure}[H]
  \centering
  \includegraphics[width=\textwidth,trim=0 92 0 22,clip]{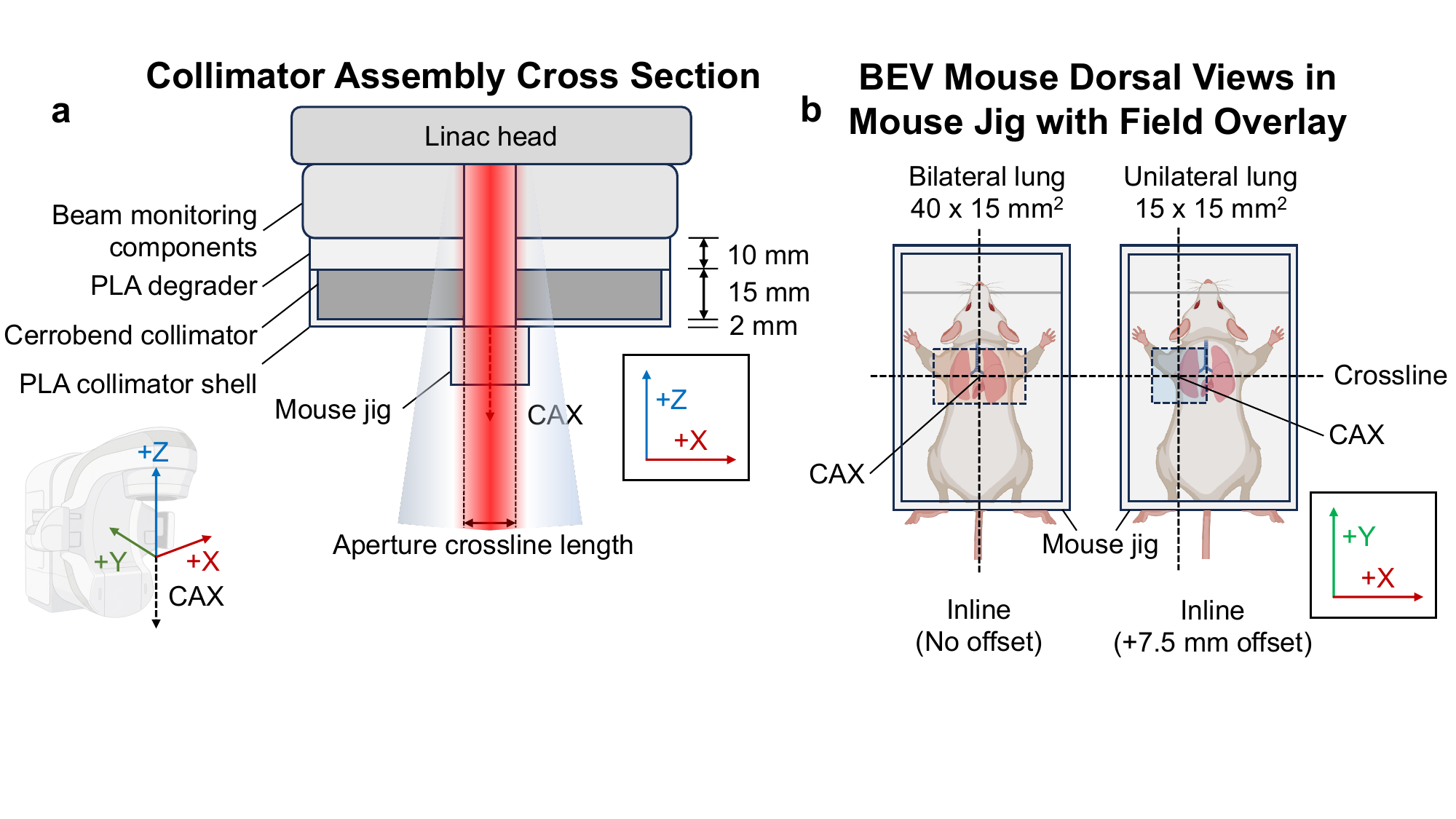}
  \caption{Linac-based preclinical setup and field layout. (\textbf{a}) Cross-sectional schematic: monitor structure, 10~mm PLA degrader, 15~mm Cerrobend collimator, 2~mm PLA collimator shell, and mouse jig; CAX and aperture crossline are indicated at specimen level. Inset: gantry-related axes. (\textbf{b}) Beam's-eye dorsal views: bilateral lung (\cm{4.0\by1.5}, CAX on midline) and unilateral lung (\cm{1.5\by1.5}, 7.5~mm inline jig offset). Axis directions match those printed on the panels.}
  \label{fig:setup}
\end{figure}

Collimator Creator (CC) generates watertight STL models for fused deposition modeling and for import into Monte Carlo geometry\cite{Perl2012,Faddegon2020}. Here, watertight denotes a closed surface mesh suitable for slicing or geometry import. The graphical interface, previewed part geometries, offset geometry relative to the collimator central axis (CAX), and modular mechanical stack are shown in \textbf{Figure~\ref{fig:cc}}. Numerical inputs map directly to dosimetric degrees of freedom in the user interface (\textbf{Figure~\ref{fig:cc}a}). Inputs include width and length for rectangular apertures, a single radius for circular apertures, or custom outline (including contour-derived shapes with margin). These are software capabilities; only the two rectangular lung fields described in Section~\ref{sec:validation} were dosimetrically evaluated. The center of the aperture aligns with the CAX by default while users can also input crossline and inline offsets in millimeters to specify the displacement from the CAX, as in \textbf{Figure~\ref{fig:cc}c}. Physically, the parametric recipe produces three functional outputs along the beam (\textbf{Figures~\ref{fig:cc}d and \ref{fig:setup}a}): (i) a 10~mm PLA beam degrader upstream of the aperture, which sets the mean electron energy and scatter incident on the block; (ii) a Cerrobend collimator (15~mm along the beam in this work) produced from a disposable PETG collimator mold. The mold is only a casting fixture, whereas the alloy defines projected field size and penumbra at the specimen. Lastly, (iii) a thin PLA collimator shell that registers the degrader and Cerrobend collimator to the custom electron applicator (and a separate PLA holder supports the demolded Cerrobend collimator). Aperture width and length (or contour-based outlines) set crossline and inline field dimensions at the specimen. The collimator shell offsets the specimen jig relative to the CAX so one block geometry can serve bilateral (centered) versus unilateral lung irradiation with a controlled lateral offset (\textbf{Figure~\ref{fig:setup}b}). The exported solids correspond to the components illustrated in \textbf{Figure~\ref{fig:cc}b}. Designs are exported as meshes for slicing, printing, and casting (Section~\ref{sec:printing}).

\subsection{3D Printing and Cerrobend Casting}
\label{sec:printing}

The STL meshes exported from CC were sliced in BambuSlicer (version 1.9.7, 2024) and fabricated on a Bambu Lab X1 Carbon FDM 3D printer (Bambu Lab, Shenzhen, China). The mold was printed in PETG. PETG was selected because its glass-transition temperature range (79--85\,\textdegree C) exceeds the melting range of Cerrobend (70--75\,\textdegree C)\cite{Yan2024}. The Cerrobend collimator holder, collimator shell, and beam degrader, which are not exposed to molten Cerrobend, were printed in standard PLA. Cerrobend was melted to 75\,\textdegree C in a thermostatically controlled pot following the block room practice recommended for electron field shaping. The molten alloy was poured in a slow, continuous stream to minimize air entrapment. Cooling proceeded in still air for a minimum of 30 minutes. After solidification, the Cerrobend collimator was demolded with taps on the aperture wall with a rubber mallet, and then fitted into its printed PLA holder, which provides mechanical support and limits operator contact with the toxic Cerrobend alloy. Kilovoltage (kV) planar images were acquired to assess collimator geometry and uniformity. Images of fabricated Cerrobend collimators were acquired with the on-board kV imaging system of the TrueBeam at a tube voltage of 50~kVp and an exposure of 1.6~mAs (40~mA, 40~ms). The length and width were measured on the image projected to the isocenter plane using the source-to-axis distance to source-to-image distance ratio. A representative image is shown in \textbf{Figure~\ref{fig:cc}f}. To assess Cerrobend defects, the collimator with the smallest aperture size (\cm{1.5\by1.5}) was selected because it contained the largest Cerrobend-covered area, thereby maximizing the area evaluated for potential defects. A region of interest (ROI) covering the central 90\% of the collimator was analyzed after excluding the aperture and a 1-mm margin around it. Uniformity was calculated as one minus the ratio of the 95th--5th percentile signal spread to the mean open-aperture signal, expressed as a percentage.

\subsection{Experimental Validation}
\label{sec:validation}

Two organ-specific collimator geometries were created for demonstration: bilateral lung collimator: \cm{4.0\by1.5} field centered on the CAX, and unilateral lung collimator: \cm{1.5\by1.5} field, offsetting the mouse jig by 7.5~mm from the beam CAX. These collimation geometries allow for small-animal irradiation used in rodent models to study radiation-induced pneumonitis and fibrosis, by either irradiating both lungs (bilateral) or left lung (unilateral). The collimator assemblies (collimator shell and Cerrobend collimator) were mounted onto a custom electron applicator attached to a Varian TrueBeam linac (\textbf{Figure~\ref{fig:setup}}). The linac configuration and platform-level beam characteristics, including percent depth dose, dose calibration, dose per pulse, and output reproducibility, were reported previously\cite{Melemenidis2026} and were not re-evaluated here. Beam-shaping performance was quantified with films, positioned flush with the downstream face of the collimator to record the in-air fluence map. Films were allowed to stabilize for 24~h, then scanned in transmission mode on an Epson Expression 10000 XL flat-bed scanner (Seiko Epson Corporation, Nagano, Japan) at 150~dpi (0.169~mm pixel spacing) with all color corrections disabled. The red-channel optical density was converted to dose using a pre-established calibration curve. One-dimensional profiles along the principal X and Y axes were extracted and analyzed with in-house MATLAB scripts. Full width at half maximum (FWHM) was calculated. The 80\%--20\% penumbra was calculated at each field edge, and the left and right values were averaged for reporting. The 80\% and 20\% crossings were obtained by interpolation between adjacent pixels. At conventional dose rate (CONV), relative doses along the crossline were measured using an ionization chamber (PTW 30010, PTW, Freiburg, Germany) positioned at the CAX and at 2~cm from the collimator edge. The measurement setup included 1~cm solid water upstream and 1~cm downstream (both simulating mouse tissue), with a 2~cm air gap between the phantom and collimator, and measurements were repeated three times at each position. The electrometer background charge was measured and subtracted from each reading, and a readout threshold was applied so that residual background signal did not contribute to the recorded charge. Off-axis percentage leakage doses were calculated relative to the CAX dose. The uncertainty of each leakage ratio was evaluated from the three repeated readings. A leakage ratio was calculated from each paired off-axis and CAX reading and the standard uncertainty was taken as the standard deviation of the three ratios. Because the leakage dose is expressed as the ratio of two readings acquired with the same chamber, electrometer, phantom, and beam configuration, the chamber calibration coefficient and the associated influence quantities cancel and do not contribute to the reported uncertainty.

\section{Results}

\subsection{Component Accuracy}

Comparisons between the CAD dimensions and the fabricated Cerrobend collimators, measured from the kV planar images, showed deviations of less than 0.2~mm for both collimator geometries. For the unilateral collimator, analysis of its central 90\% ROI yielded a uniformity of 99.1\%, with no pixels exceeding 5\% of the mean open-aperture signal, indicating no observable Cerrobend defects. Removal from the PETG molds was completed with the use of a rubber mallet. When assembled in the printed enclosing case, the Cerrobend collimator fit snugly with minimal post-processing.

\subsection{Beam Profile Measurements}

\begin{figure}[H]
  \centering
  \includegraphics[width=0.8\textwidth]{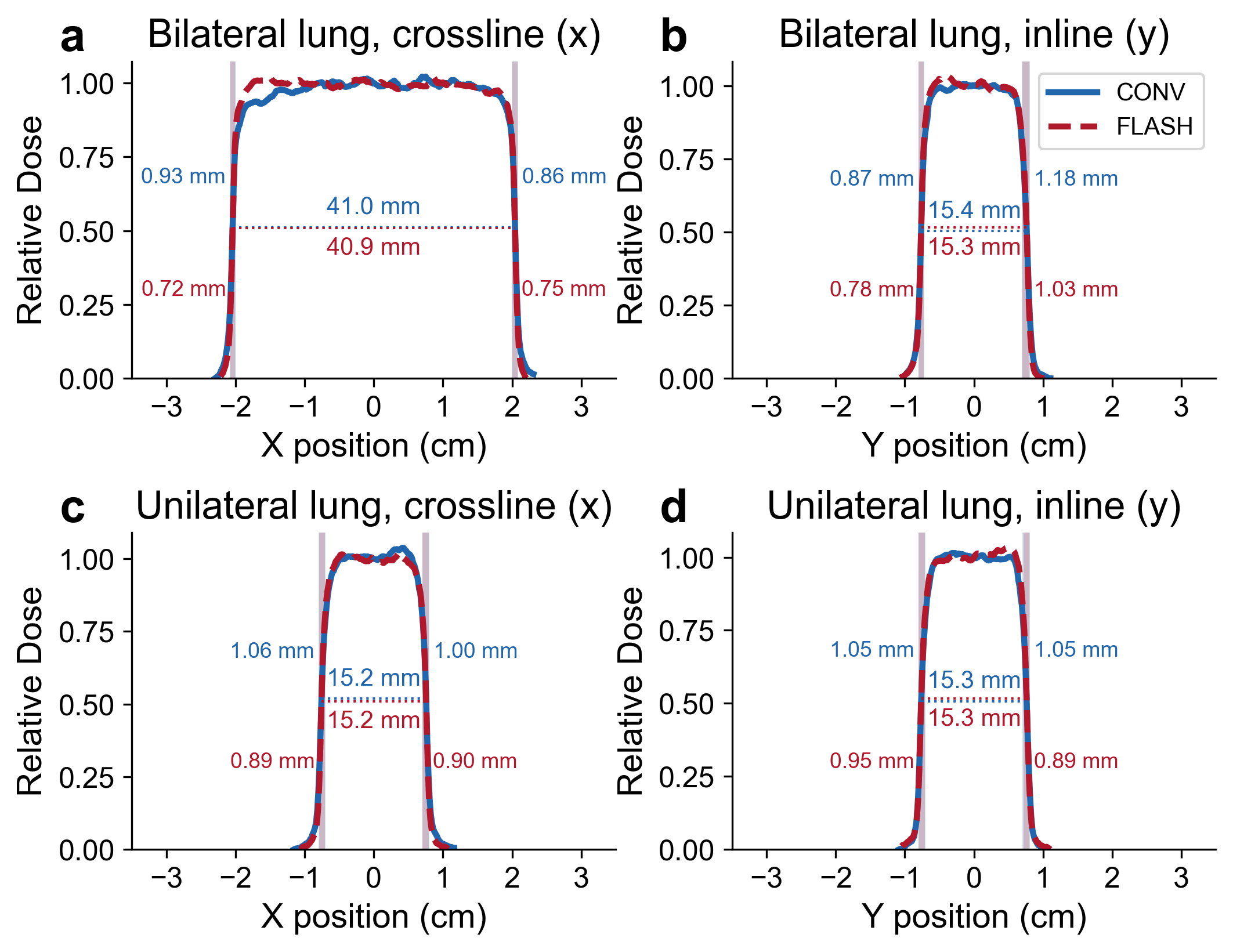}
  \caption{Radiochromic film profiles at the downstream face of the collimator for conventional dose rate (CONV, solid blue) and UHDR (FLASH, dashed red). (\textbf{a}) Bilateral lung, crossline (x). (\textbf{b}) Bilateral lung, inline (y). (\textbf{c}) Unilateral lung, crossline (x). (\textbf{d}) Unilateral lung, inline (y). Annotated values in (\textbf{a--d}) are the full width at half maximum (FWHM) at 50\% of central-axis dose and the 80\%--20\% penumbra, both in millimeters.}
  \label{fig:profiles}
\end{figure}

Film dosimetry confirmed that collimators for both bilateral lung and unilateral lung irradiations produced beam profiles that closely aligned with the designed field dimensions when placed directly on the underside of the PLA collimator shell (\textbf{Figure~\ref{fig:profiles}}). For the bilateral lung collimator (\cm{4.0\by1.5}), at the CONV dose rate, the measured FWHM was 4.10~cm in the x-axis and 1.54~cm in the y-axis, while under UHDR conditions, the equivalent FWHM values were 4.09~cm and 1.53~cm, respectively. At the CONV dose rate, the measured mean 80\%--20\% penumbra was 0.89~mm in the x-axis and 1.03~mm in the y-axis, while under UHDR conditions, the equivalent penumbra values were 0.73~mm and 0.90~mm, respectively. The relative leakage dose along crossline measured using the ionization chamber at 4~cm from the central axis, i.e.\ 2~cm from the collimator edge, was $1.6226 \pm 0.0002$\%. Similarly, for the unilateral lung collimator with field size (\cm{1.5\by1.5}), at the CONV dose rate, the measured FWHM was 1.52~cm in the x-axis and 1.53~cm in y-axis, while for the UHDR conditions the equivalent FWHM values were 1.52~cm and 1.53~cm, respectively. At the CONV dose rate, the measured mean 80\%--20\% penumbra was 1.03~mm in the x-axis and 1.05~mm in the y-axis, while for the UHDR conditions the equivalent penumbra values were 0.90~mm and 0.92~mm, respectively. All the FWHM values are within 3\% difference from their designed widths. The relative leakage dose measured using the ionization chamber at 3.5~cm from the central axis, i.e.\ 2~cm from the collimator edge, was $1.3102 \pm 0.0002$\%.

\section{Discussion}

\subsection{Reproducibility and Applications in Preclinical Irradiation}

Traditional beam shaping with hand-crafted Cerrobend blocks can introduce substantial variability in field shape and position, as casting and cutting are prone to human error. In one study comparing electron beam cutouts, manually fabricated Cerrobend inserts deviated from planned field outlines by 1--3~mm, whereas 3D printed alternatives achieved alignment within 0.4~mm\cite{Skinner2019}. The present study did not directly compare fabrication time, inter-user variability, or manufacturing reproducibility with a manual workflow. This variability in block production can lead to inconsistent dosing and treatment effects in preclinical studies, exacerbating the well-documented challenges in standardizing radiobiology experiments across different labs. Indeed, a multi-center audit of preclinical irradiation showed that lack of standardized dosimetry protocols contributed to dose discrepancies over 10\% between institutions\cite{SilvestrePatallo2020}. In contrast, CC defines collimator geometries programmatically, so identical STL models can be produced and shared. This digital reproducibility addresses calls for greater consistency in preclinical radiotherapy techniques. For example, Ashraf et al.\ reported that attaining dosimetric consensus across facilities is critical to reproducibility in UHDR animal experiments\cite{Ashraf2024}. This work contributes to this goal by replacing manually fabricated blocks and designs with algorithmically generated designs that can be reproduced from the same inputs. By standardizing digital design inputs, CC may support more consistent \textit{in vivo} studies; reproducibility of the complete fabrication and dosimetric process was not evaluated.

Beyond improving reproducibility, the tool may also enhance the precision of single-beam radiotherapy applications of small-animal irradiators. Some linac-based preclinical electron studies rely on single-beam open fields or basic collimators, which limits the ability to conform dose to targets or spare normal tissues. Our system enables customized fields (e.g.\ apertures for brain, thorax, or tumor-specific shapes) to be rapidly generated, lowering the barrier to conformal techniques in laboratories without dedicated small animal radiosurgery platforms. Standardizing such beam shapes can also help harmonize research protocols. Although this work demonstrates rectangular fields, the custom JSON functionality can generate contours derived from CT projections or other 2D contouring methods, whose geometries were not dosimetrically validated here. Recent consensus recommendations in radiation biology have highlighted that a lack of standardization in preclinical studies hinders cross-comparison\cite{Biglin2022}. A library of validated collimator designs could support multi-institutional use of the same beam geometries, although design transfer across institutions was not evaluated in this study.

The code is made available at \url{https://github.com/wizrads/CollimatorCreator} with detailed documentation.

\subsection{Parametric CAD and 3D Printing for Electron Cutout Workflows}

The clinical electron cutout workflow for superficial lesions provides a direct analogue for CC: 3D printing can standardize field-shaping geometry while retaining established electron-cutout fabrication steps\cite{Skinner2019,Schulz2023}. For preclinical applications, automated STL generation by CC enables a similar workflow for custom linac electron collimators. A digital design can be adjusted with a few parameter changes and printed without reconstructing the CAD model manually, although fabrication time was not measured in this study. The integration of CAD reduces repeated manual geometry construction. Prior preclinical studies have demonstrated 3D-printed compensator molds and blocks\cite{Redler2021,Yoon2020}. These examples support the feasibility of printed beam-shaping accessories. Another benefit is the ease of sharing and scaling designs. A digital STL model can be distributed electronically, so researchers at different sites can print the same accessory. Byrne et al.\ described a platform-specific linac UHDR workflow using Cerrobend collimation\cite{Byrne2024}. CC's distinct contribution is open-source parametric generation of STL components with integrated aperture and jig offsets. In line with modern open-science practices, a repository of validated collimator designs for common preclinical scenarios could support standardized geometry.

\subsection{UHDR Radiotherapy-Specific Considerations}

The present measurements compare CC-generated field widths under conventional and UHDR delivery but do not test a dose-rate effect. Early \textit{in vivo} UHDR studies were constrained by fixed beam geometries; for example, modified linacs used for electron UHDR lacked any mechanism to adjust the beam shape or size, resulting in suboptimal field coverage\cite{DiMartino2020}. Investigators noted that without flexible collimation, it was impossible to accurately irradiate well-defined volumes or to study how the UHDR effect varies with field size. CC's ability to produce custom collimators addresses this by enabling precise definition of the irradiation field. For example, a recent report of a murine UHDR whole-thorax irradiation platform utilized a \cm{15\by1.6} Cerrobend slit to concurrently irradiate multiple mice and achieved highly uniform dose distributions up to 120~Gy/s average dose rate\cite{Byrne2024}. The authors emphasized that accurately collimated, wide uniform fields enabled simultaneous treatment of three animals with consistent dosing, a setup needed for high-throughput UHDR studies. This illustrates how custom-designed collimators can be integral to UHDR research infrastructure. The current study evaluates CC-generated field shaping only and does not assess the biological FLASH effect. Platform-level beam characterization and UHDR dosimetry were reported previously\cite{Melemenidis2026} and were not re-evaluated here.

\subsection{Limitations and Future Directions}

While the presented software improves upon the status quo in several ways, the current validation is limited to two rectangular fields measured with film at the downstream face of the collimator and it does not establish output factors, surface-dose perturbation, or dose at the specimen location. Circular and custom apertures were not dosimetrically validated, and fabrication time, inter-user variability, and manufacturing reproducibility were not compared with a manual workflow. A potential future enhancement is integration with image guidance and treatment planning. Currently, the user specifies dimensions or uploads contours manually and automatic generation from a delineated target was not tested. CC exports standard STL files that could be imported into Monte Carlo radiation transport simulations\cite{Perl2012,Faddegon2020}. No Monte Carlo simulation, simulation-to-measurement comparison, or inverse design was performed in this study. Monte Carlo integration and optimization are therefore future directions rather than validated components of the present workflow. Future work could compare simulated and measured dose for CC-generated geometries before evaluating optimization methods such as TopasOpt\cite{Whelan2023}. Another limitation is that the current tool produces static apertures of uniform thickness. Future versions could evaluate 3D-printed compensators or variable-thickness collimators, but these capabilities require separate dosimetric validation\cite{Redler2021}.

\section{Conclusions}

Collimator Creator provides an open-source parametric workflow that converts user-defined aperture dimensions and jig offsets into printable components for linac-based preclinical electron collimation. For two rectangular lung-field designs, the fabricated Cerrobend apertures agreed with the specified dimensions within 0.2~mm, and the measured field widths were within 3\% of the nominal dimensions under conventional and UHDR delivery. The 80\%--20\% penumbra ranged from 0.73 to 1.05~mm, while conventional-dose-rate leakage measured 2~cm beyond the collimator edge was below 2\% for both designs. These results demonstrate accurate translation of parametric digital designs into field-shaping hardware for the evaluated configurations. Further work will assess specimen-level dosimetry, additional aperture geometries, and Monte Carlo integration.

\section*{Funding}

This research received no specific grant from any funding agency in the public, commercial, or not-for-profit sectors.

\section*{Data Availability}

The code supporting the software introduced in this study is openly available at \url{https://github.com/wizrads/CollimatorCreator}.

\bibliographystyle{medphys-ama}
\bibliography{refs}

\begin{thebibliography}{34}
\providecommand{\natexlab}[1]{#1}
\providecommand{\url}[1]{\texttt{#1}}
\expandafter\ifx\csname urlstyle\endcsname\relax
  \providecommand{\doi}[1]{doi: #1}\else
  \providecommand{\doi}{doi: \begingroup \urlstyle{rm}\Url}\fi

\bibitem[Hall and Giaccia(2006)]{Hall2006}
Hall EJ, Giaccia AJ.
\newblock \emph{Radiobiology for the Radiologist}.
\newblock Lippincott Williams \& Wilkins, Philadelphia, 6th edition, 2006.

\bibitem[Kirsch et~al.(2018)Kirsch, Diehn, Kesarwala, et~al.]{Kirsch2018}
Kirsch DG, Diehn M, Kesarwala AH, et~al.
\newblock The future of radiobiology.
\newblock \emph{J Natl Cancer Inst}, 110\penalty0 (4):\penalty0 329--340, 2018.
\newblock \doi{10.1093/jnci/djx231}.

\bibitem[Verhaegen et~al.(2011)Verhaegen, Granton, and
  Tryggestad]{Verhaegen2011}
Verhaegen F, Granton P, Tryggestad E.
\newblock Small animal radiotherapy research platforms.
\newblock \emph{Phys Med Biol}, 56\penalty0 (12):\penalty0 R55, 2011.
\newblock \doi{10.1088/0031-9155/56/12/R01}.

\bibitem[Brown et~al.(2022)Brown, Ghita, Dubois, et~al.]{Brown2022}
Brown KH, Ghita M, Dubois LJ, et~al.
\newblock A scoping review of small animal image-guided radiotherapy research:
  Advances, impact and future opportunities in translational radiobiology.
\newblock \emph{Clin Transl Radiat Oncol}, 34:\penalty0 112--119, 2022.
\newblock \doi{10.1016/j.ctro.2022.04.004}.

\bibitem[Pizarro and Hern{\'a}ndez(2017)]{Pizarro2017}
Pizarro F, Hern{\'a}ndez A.
\newblock Optimization of radiotherapy fractionation schedules based on
  radiobiological functions.
\newblock \emph{Br J Radiol}, 90\penalty0 (1079):\penalty0 20170400, 2017.
\newblock \doi{10.1259/bjr.20170400}.

\bibitem[Sch{\"u}ler et~al.(2022)Sch{\"u}ler, Acharya, Montay-Gruel, {Loo Jr},
  Vozenin, and Maxim]{Schuler2022}
Sch{\"u}ler E, Acharya M, Montay-Gruel P, {Loo Jr} BW, Vozenin MC, Maxim PG.
\newblock Ultra-high dose rate electron beams and the {FLASH} effect: From
  preclinical evidence to a new radiotherapy paradigm.
\newblock \emph{Med Phys}, 49\penalty0 (3):\penalty0 2082--2095, 2022.
\newblock \doi{10.1002/mp.15442}.

\bibitem[Bazalova et~al.(2014)Bazalova, Nelson, Noll, and Graves]{Bazalova2014}
Bazalova M, Nelson G, Noll JM, Graves EE.
\newblock Modality comparison for small animal radiotherapy: A simulation
  study.
\newblock \emph{Med Phys}, 41\penalty0 (1):\penalty0 011710, 2014.
\newblock \doi{10.1118/1.4842415}.

\bibitem[Graves et~al.(2007)Graves, Zhou, Chatterjee, et~al.]{Graves2007}
Graves EE, Zhou H, Chatterjee R, et~al.
\newblock Design and evaluation of a variable aperture collimator for conformal
  radiotherapy of small animals using a {microCT} scanner.
\newblock \emph{Med Phys}, 34\penalty0 (11):\penalty0 4359--4367, 2007.
\newblock \doi{10.1118/1.2789498}.

\bibitem[Wong et~al.(2008)Wong, Armour, Kazanzides, et~al.]{Wong2008}
Wong J, Armour E, Kazanzides P, et~al.
\newblock A high resolution small animal radiation research platform ({SARRP})
  with x-ray tomographic guidance capabilities.
\newblock \emph{Int J Radiat Oncol Biol Phys}, 71\penalty0 (5):\penalty0
  1591--1599, 2008.
\newblock \doi{10.1016/j.ijrobp.2008.04.025}.

\bibitem[Sch{\"u}ler et~al.(2017)Sch{\"u}ler, Trovati, King,
  et~al.]{Schuler2017}
Sch{\"u}ler E, Trovati S, King G, et~al.
\newblock Experimental platform for ultra-high dose rate {FLASH} irradiation of
  small animals using a clinical linear accelerator.
\newblock \emph{Int J Radiat Oncol Biol Phys}, 97\penalty0 (1):\penalty0
  195--203, 2017.
\newblock \doi{10.1016/j.ijrobp.2016.09.018}.

\bibitem[Lempart et~al.(2019)Lempart, Blad, Adrian, et~al.]{Lempart2019}
Lempart M, Blad B, Adrian G, et~al.
\newblock Modifying a clinical linear accelerator for delivery of ultra-high
  dose rate irradiation.
\newblock \emph{Radiother Oncol}, 139:\penalty0 40--45, 2019.
\newblock \doi{10.1016/j.radonc.2019.01.031}.

\bibitem[Jaccard et~al.(2018)Jaccard, Dur{\'a}n, Petersson,
  et~al.]{Jaccard2018}
Jaccard M, Dur{\'a}n MT, Petersson K, et~al.
\newblock High dose-per-pulse electron beam dosimetry: Commissioning of the
  {Oriatron} {eRT6} prototype linear accelerator for preclinical use.
\newblock \emph{Med Phys}, 45\penalty0 (2):\penalty0 863--874, 2018.
\newblock \doi{10.1002/mp.12713}.

\bibitem[Verhaegen et~al.(2018)Verhaegen, Dubois, Gianolini,
  et~al.]{Verhaegen2018}
Verhaegen F, Dubois L, Gianolini S, et~al.
\newblock {ESTRO ACROP}: Technology for precision small animal radiotherapy
  research: Optimal use and challenges.
\newblock \emph{Radiother Oncol}, 126\penalty0 (3):\penalty0 471--478, 2018.
\newblock \doi{10.1016/j.radonc.2017.11.016}.

\bibitem[Hackam and Redelmeier(2006)]{Hackam2006}
Hackam DG, Redelmeier DA.
\newblock Translation of research evidence from animals to humans.
\newblock \emph{JAMA}, 296\penalty0 (14):\penalty0 1731--1732, 2006.
\newblock \doi{10.1001/jama.296.14.1731}.

\bibitem[Redler et~al.(2021)Redler, Pearson, Liu, et~al.]{Redler2021}
Redler G, Pearson E, Liu X, et~al.
\newblock Small animal {IMRT} using {3D}-printed compensators.
\newblock \emph{Int J Radiat Oncol Biol Phys}, 110\penalty0 (2):\penalty0
  551--565, 2021.
\newblock \doi{10.1016/j.ijrobp.2020.12.028}.

\bibitem[Byrne et~al.(2024)Byrne, Poirier, Xu, et~al.]{Byrne2024}
Byrne KE, Poirier Y, Xu J, et~al.
\newblock Technical note: A small animal irradiation platform for investigating
  the dependence of the {FLASH} effect on electron beam parameters.
\newblock \emph{Med Phys}, 51\penalty0 (2):\penalty0 1421--1432, 2024.
\newblock \doi{10.1002/mp.16909}.

\bibitem[No et~al.(2023)No, Wu, Dworkin, et~al.]{No2023}
No HJ, Wu Y, Dworkin ML, et~al.
\newblock Clinical linear accelerator-based electron {FLASH}: Pathway for
  practical translation to {FLASH} clinical trials.
\newblock \emph{Int J Radiat Oncol Biol Phys}, 117\penalty0 (2):\penalty0
  482--492, 2023.
\newblock \doi{10.1016/j.ijrobp.2023.04.011}.

\bibitem[Draeger et~al.(2020)Draeger, Sawant, Johnstone, et~al.]{Draeger2020}
Draeger E, Sawant A, Johnstone C, et~al.
\newblock A dose of reality: How 20 years of incomplete physics and dosimetry
  reporting in radiobiology studies may have contributed to the reproducibility
  crisis.
\newblock \emph{Int J Radiat Oncol Biol Phys}, 106\penalty0 (2):\penalty0
  243--252, 2020.
\newblock \doi{10.1016/j.ijrobp.2019.06.2545}.

\bibitem[Rooney et~al.(2020)Rooney, Rosenberg, Braunstein, et~al.]{Rooney2020}
Rooney MK, Rosenberg DM, Braunstein S, et~al.
\newblock Three-dimensional printing in radiation oncology: A systematic review
  of the literature.
\newblock \emph{J Appl Clin Med Phys}, 21\penalty0 (8):\penalty0 15--26, 2020.
\newblock \doi{10.1002/acm2.12907}.

\bibitem[McGarry et~al.(2025)McGarry, {Tonino Baldion}, Burnley,
  et~al.]{McGarry2025}
McGarry CK, {Tonino Baldion} A, Burnley J, et~al.
\newblock {IPEM} topical report: guidance on {3D} printing in radiotherapy.
\newblock \emph{Phys Med Biol}, 70\penalty0 (4):\penalty0 04TR01, 2025.
\newblock \doi{10.1088/1361-6560/ada518}.

\bibitem[Skinner et~al.(2019)Skinner, Fahimian, and Yu]{Skinner2019}
Skinner L, Fahimian BP, Yu AS.
\newblock Tungsten filled {3D} printed field shaping devices for electron beam
  radiation therapy.
\newblock \emph{PLOS ONE}, 14\penalty0 (6):\penalty0 e0217757, 2019.
\newblock \doi{10.1371/journal.pone.0217757}.

\bibitem[Schulz et~al.(2023)Schulz, Gibson, Dubrowski, et~al.]{Schulz2023}
Schulz JB, Gibson C, Dubrowski P, et~al.
\newblock Shaping success: clinical implementation of a {3D}-printed electron
  cutout program in external beam radiation therapy.
\newblock \emph{Front Oncol}, 13:\penalty0 1237037, 2023.
\newblock \doi{10.3389/fonc.2023.1237037}.

\bibitem[Schulz et~al.(2024)Schulz, Dubrowski, Gibson, Yu, and
  Skinner]{Schulz2024}
Schulz JB, Dubrowski P, Gibson C, Yu AS, Skinner LB.
\newblock A clinical solution for non-toxic {3D}-printed photon blocks in
  external beam radiation therapy.
\newblock \emph{J Appl Clin Med Phys}, 25\penalty0 (3):\penalty0 e14225, 2024.
\newblock \doi{10.1002/acm2.14225}.

\bibitem[Capaldi et~al.(2024)Capaldi, Gibson, Villa, et~al.]{Capaldi2024}
Capaldi DPI, Gibson C, Villa A, et~al.
\newblock Tungsten filled 3-dimensional printed lung blocks for total body
  irradiation.
\newblock \emph{Pract Radiat Oncol}, 14\penalty0 (3):\penalty0 267--276, 2024.
\newblock \doi{10.1016/j.prro.2023.11.003}.

\bibitem[Melemenidis et~al.(2026)Melemenidis, Chen, Jensen,
  et~al.]{Melemenidis2026}
Melemenidis S, Chen D, Jensen C, et~al.
\newblock Noninvasive reversible software-based electron {FLASH} irradiation
  configuration of a linear accelerator in clinical use.
\newblock \emph{Int J Radiat Oncol Biol Phys}, 2026.
\newblock \doi{10.1016/j.ijrobp.2026.05.010}.

\bibitem[Perl et~al.(2012)Perl, Shin, Schumann, Faddegon, and
  Paganetti]{Perl2012}
Perl J, Shin J, Schumann J, Faddegon B, Paganetti H.
\newblock {TOPAS}: An innovative proton {Monte Carlo} platform for research and
  clinical applications.
\newblock \emph{Med Phys}, 39:\penalty0 6818, 2012.
\newblock \doi{10.1118/1.4758060}.

\bibitem[Faddegon et~al.(2020)Faddegon, Ramos-Mendez, Schuemann,
  et~al.]{Faddegon2020}
Faddegon B, Ramos-Mendez J, Schuemann J, et~al.
\newblock The {TOPAS} tool for particle simulation, a {Monte Carlo} simulation
  tool for physics, biology and clinical research.
\newblock \emph{Phys Med}, 72:\penalty0 114--121, 2020.
\newblock \doi{10.1016/j.ejmp.2020.03.019}.

\bibitem[Yan et~al.(2024)Yan, Kleiner, Tabigue, et~al.]{Yan2024}
Yan C, Kleiner C, Tabigue A, et~al.
\newblock {PETG}: Applications in modern medicine.
\newblock \emph{Eng Regen}, 5\penalty0 (1):\penalty0 45--55, 2024.
\newblock \doi{10.1016/j.engreg.2023.11.001}.

\bibitem[{Silvestre Patallo} et~al.(2020){Silvestre Patallo}, Subiel,
  Westhorpe, et~al.]{SilvestrePatallo2020}
{Silvestre Patallo} I, Subiel A, Westhorpe A, et~al.
\newblock Development and implementation of an end-to-end test for absolute
  dose verification of small animal preclinical irradiation research platforms.
\newblock \emph{Int J Radiat Oncol Biol Phys}, 107\penalty0 (3):\penalty0
  587--596, 2020.
\newblock \doi{10.1016/j.ijrobp.2020.03.001}.

\bibitem[Ashraf et~al.(2024)Ashraf, Melemenidis, Liu, et~al.]{Ashraf2024}
Ashraf MR, Melemenidis S, Liu K, et~al.
\newblock Multi-institutional audit of {FLASH} and conventional dosimetry with
  a {3D} printed anatomically realistic mouse phantom.
\newblock \emph{Int J Radiat Oncol Biol Phys}, 120\penalty0 (1):\penalty0
  287--300, 2024.
\newblock \doi{10.1016/j.ijrobp.2024.03.017}.

\bibitem[Biglin et~al.(2022)Biglin, Aitkenhead, Price, et~al.]{Biglin2022}
Biglin ER, Aitkenhead AH, Price GJ, et~al.
\newblock A preclinical radiotherapy dosimetry audit using a realistic {3D}
  printed murine phantom.
\newblock \emph{Sci Rep}, 12\penalty0 (1):\penalty0 6826, 2022.
\newblock \doi{10.1038/s41598-022-10895-5}.

\bibitem[Yoon et~al.(2020)Yoon, Kodra, Miles, Kirsch, and Oldham]{Yoon2020}
Yoon SW, Kodra J, Miles DA, Kirsch DG, Oldham M.
\newblock A method for generating intensity-modulated radiation therapy fields
  for small animal irradiators utilizing {3D}-printed compensator molds.
\newblock \emph{Med Phys}, 47\penalty0 (9):\penalty0 4363--4371, 2020.
\newblock \doi{10.1002/mp.14175}.

\bibitem[{Di Martino} et~al.(2020){Di Martino}, Barca, Barone,
  et~al.]{DiMartino2020}
{Di Martino} F, Barca P, Barone S, et~al.
\newblock {FLASH} radiotherapy with electrons: Issues related to the
  production, monitoring, and dosimetric characterization of the beam.
\newblock \emph{Front Phys}, 8:\penalty0 570697, 2020.
\newblock \doi{10.3389/fphy.2020.570697}.

\bibitem[Whelan et~al.(2023)Whelan, {Loo Jr}, Wang, and Keall]{Whelan2023}
Whelan B, {Loo Jr} BW, Wang J, Keall P.
\newblock {TopasOpt}: An open-source library for optimization with {Topas}
  {Monte Carlo}.
\newblock \emph{Med Phys}, 50\penalty0 (2):\penalty0 1121--1131, 2023.
\newblock \doi{10.1002/mp.16126}.

\end{thebibliography}

\end{document}